\documentclass[11pt,notitlepage,superscriptaddress,floatfix,prl]{revtex4-2}

\newcommand{\nm}{\bar{n}}

\newcommand{\Gm}{\Gamma_{\mathrm{m}}}
\newcommand{\Gmj}{\Gamma_{\mathrm{m}j}}
\newcommand{\GmX}{\Gamma_{\mathrm{m}\scriptscriptstyle{X}}}
\newcommand{\GmY}{\Gamma_{\mathrm{m}\scriptscriptstyle{Y}}}
\newcommand{\Gnj}{\Gamma_{\mathrm{n}j}}
\newcommand{\GnX}{\Gamma_{\mathrm{n}\scriptscriptstyle{X}}}
\newcommand{\GnY}{\Gamma_{\mathrm{n}\scriptscriptstyle{Y}}}

\newcommand{\Omegamj}{\Omega_{j}}
\newcommand{\OmegaX}{\Omega_{\scriptscriptstyle{X}}^0}
\newcommand{\OmegaY}{\Omega_{\scriptscriptstyle{Y}}^0}
\newcommand{\gX}{g_{\scriptscriptstyle{X}}}
\newcommand{\gY}{g_{\scriptscriptstyle{Y}}}
\newcommand{\bX}{\hat{b}_{\scriptscriptstyle{X}}}
\newcommand{\bY}{\hat{b}_{\scriptscriptstyle{Y}}}
\newcommand{\bdX}{\hat{b}^{\dagger}_{\scriptscriptstyle{X}}}
\newcommand{\bdY}{\hat{b}^{\dagger}_{\scriptscriptstyle{Y}}}
\newcommand{\tbX}{\tilde{b}_{\scriptscriptstyle{X}}}
\newcommand{\tbY}{\tilde{b}_{\scriptscriptstyle{Y}}}
\newcommand{\tbdX}{\tilde{b}^{\dagger}_{\scriptscriptstyle{X}}}
\newcommand{\tbdY}{\tilde{b}^{\dagger}_{\scriptscriptstyle{Y}}}

\newcommand{\ac}{\hat{a}_{\mathrm{c}}}
\newcommand{\tac}{\tilde{a}_{\mathrm{c}}}

\newcommand{\kB}{k_{\mathrm{B}}}

\newcommand{\ud}{\mathrm{d}}

\usepackage{amsmath}
\usepackage[dvips]{graphicx}
\usepackage{epsfig}
\usepackage{tabularx}
\usepackage{color}
\usepackage[T1]{fontenc}

\begin{document}

\title{Vectorial polaritons in the quantum motion of a levitated nanosphere}

\author{A. Ranfagni}
\affiliation{European Laboratory for Non-Linear Spectroscopy (LENS), via Carrara 1, I-50019 Sesto Fiorentino (FI), Italy}
\affiliation{INFN, Sezione di Firenze, via Sansone 1, I-50019 Sesto Fiorentino (FI), Italy}

\author{P. Vezio}
\affiliation{European Laboratory for Non-Linear Spectroscopy (LENS), via Carrara 1, I-50019 Sesto Fiorentino (FI), Italy}
\affiliation{INFN, Sezione di Firenze, via Sansone 1, I-50019 Sesto Fiorentino (FI), Italy}

\author{M. Calamai}
\affiliation{INFN, Sezione di Firenze, via Sansone 1, I-50019 Sesto Fiorentino (FI), Italy}

\author{A. Chowdhury\footnote{Present address: Department of Physics, University of Konstanz, 78457 Konstanz, Germany}} 
\affiliation{CNR-INO, largo Enrico Fermi 6, I-50125 Firenze, Italy}

\author{F. Marino}
\affiliation{CNR-INO, largo Enrico Fermi 6, I-50125 Firenze, Italy}
\affiliation{INFN, Sezione di Firenze, via Sansone 1, I-50019 Sesto Fiorentino (FI), Italy}

\author{F. Marin}
\email[Electronic mail: ]{marin@fi.infn.it}
\affiliation{Dipartimento di Fisica e Astronomia, Universit\`a di Firenze, via Sansone 1, I-50019 Sesto Fiorentino (FI), Italy}
\affiliation{European Laboratory for Non-Linear Spectroscopy (LENS), via Carrara 1, I-50019 Sesto Fiorentino (FI), Italy}
\affiliation{INFN, Sezione di Firenze, via Sansone 1, I-50019 Sesto Fiorentino (FI), Italy}
\affiliation{CNR-INO, largo Enrico Fermi 6, I-50125 Firenze, Italy}

\date{\today}

\maketitle

The strong coupling between elementary excitations of the electromagnetic field (photons) and quantized mechanical vibrations (phonons) produces hybrid quasi-particle states, known as phonon-polaritons \cite{tolpygo,huang,hopfield}. Their typical signature is the avoided crossing between the eigenfrequencies of the coupled system, as paradigmatically illustrated by the Jaynes-Cummings Hamiltonian \cite{jc}, and observed in quantum electrodynamics experiments where cavity photons are coupled to atoms \cite{thompson,colombe}, ions \cite{nakamura}, excitons \cite{weisbuch,reithmaier,yoshie}, spin ensambles \cite{huebl,tabuchi} and superconducting qubits \cite{wallraff}. In this work, we demonstrate the generation of phonon-polaritons in the quantum motion of an optically-levitated nanosphere \cite{Ashkin:1970,Chang:2010,Barker:2010,Romero:2010,Li:2011,Gieseler:2012,Millen:2020}. The particle is trapped in high vacuum by an optical tweezer and strongly coupled to a single cavity mode by coherent scattering of the tweezer photons \cite{delic2019,windey2019,gonzalez-ballestrero2019,delic2020,quidant2020}. The two-dimensional motion splits into two nearly-degenerate components that, together with the optical cavity mode, define an optomechanical system with three degrees-of-freedom. As such, when entering the strong coupling regime, we observe hybrid light-mechanical states with a dispersion law typical of tripartite quantum systems \cite{sun,altomare,fink}. Remarkably, the independent components of motion here identify a physical vibration direction on a plane that, similarly to the polarization of light, confers a vectorial nature to the polariton field. Our results pave the way to novel protocols for quantum information transfer between photonic and phononic components and represent a key-step towards the demonstration of optomechanical entangled states at room temperature.

The optomechanical coupling between a mechanical resonator and a cavity mode of the electromagnetic field leads to the thermalization of the former toward the temperature of the photonic bath. For weak coupling, the system dynamics can be effectively described in terms of a mechanical oscillator at angular frequency $\Omega_{\mathrm{m}}$ and an optical oscillator at frequency $-\Delta = -(\omega_{\mathrm{L}} - \omega_{\mathrm{c}})$, given by the detuning between the laser angular frequency $\omega_{\mathrm{L}}$ and the cavity resonance $\omega_{\mathrm{c}}$. Each one of these oscillators is characterized by its own damping rate: the mechanical damping $\Gamma_{\mathrm{m}}$ for the former, and the cavity decay rate $\kappa$ for the latter.  

In the strong-coupling regime, i.e., when the coupling rate $g$ exceeds the damping (namely, $4g \geq \Gamma_{\mathrm{m}}, \,\kappa$), the optical and mechanical modes can no longer be treated as separate entities and form hybrid optomechanical states \cite{groblacher,teufel-li,quidant2020,verhagen,teufel2011,wollman,pirkka,marquardt2007}. The eigenfrequencies of the system, that without any coupling would be degenerate at $\Delta = -\Omega_{\mathrm{m}}$, undergo avoided crossing with a maximum separation at resonance equal to the vacuum Rabi splitting $2 g$. As the cavity detuning is varied, they define a dispersion relation consisting of two branches with complementary asymptotic behavior: photon-like at low detuning and phonon-like at high detuning for the upper branch, and viceversa for the lower one. 

When the optomechanical coupling is larger than the phononic thermalization rate ($2g > \Gamma_{\mathrm{th}}$), the swapping time, i.e. the time at which individual photons and phonons exchange their energy, becomes shorter than the decoherence time. In this case quantum-coherent coupling between optical and mechanical modes is achieved, and the collective excitations of the system can be meaningfully described in terms of quantized polariton states. This regime has been reached so far in whispering gallery mode resonators \cite{verhagen} and cavity electromechanical systems \cite{teufel2011,wollman,pirkka}.
 
In our experiment, phonon-polaritons are shown to emerge from the quantum-coherent coupling between a cavity mode and the two-dimensional (2D) motion of an optically-levitated nanosphere. The particle dynamics occurs on the plane orthogonal to the tweezer axis and can be decomposed into two independent mechanical modes oscillating along perpendicular directions defined by the linear polarization of the tweezer field. A particular mixture of the two modes thus identifies the orientation of the Cartesian axes used to describe the particle motion. Owing to the optical spring effect, the oscillation frequencies are shifted towards degeneracy and the two oscillators become correlated even for weak optomechanical coupling. In this case, the system can be conveniently represented in terms of alternative states, given by linear combinations of the original mechanical modes, that are respectively well coupled (bright mode) and weakly coupled (dark mode) to the intracavity field \cite{genes2008,massel2012,shkarin}. As we shall see, when achieving the strong coupling regime, such states give rise to a polaritonic dispersion relation characterized by two avoided crossings centred at different frequencies. Similar dispersion curves, which provide an unambiguous evidence of strong three-body interactions \cite{sun,altomare,fink}, had never been observed in optomechanical experiments.
 
In most quantum electrodynamics and optomechanical systems, the polarization of the cavity field plays a marginal role and polaritons are typically considered as scalar bosons. While polarization effects can be ignored even in our setup, the picture is deeply modified due to the 2D dynamics of the particle. We remark indeed that here the two mechanical modes represent the orthogonal components of the particle motion. In contrast with most platforms, their linear superposition has therefore a clear physical meaning, representing a position vector that can be associated to a physical vibration direction on a plane. Due to the degeneracy of mechanical frequencies and the consequent strong correlation of modes, the 2D motion confers a kind of directionality to the associated phonons and, in the strong-coupling regime, a peculiar vectorial nature to the polariton field. 

\begin{figure}
\includegraphics[width=0.8\columnwidth]{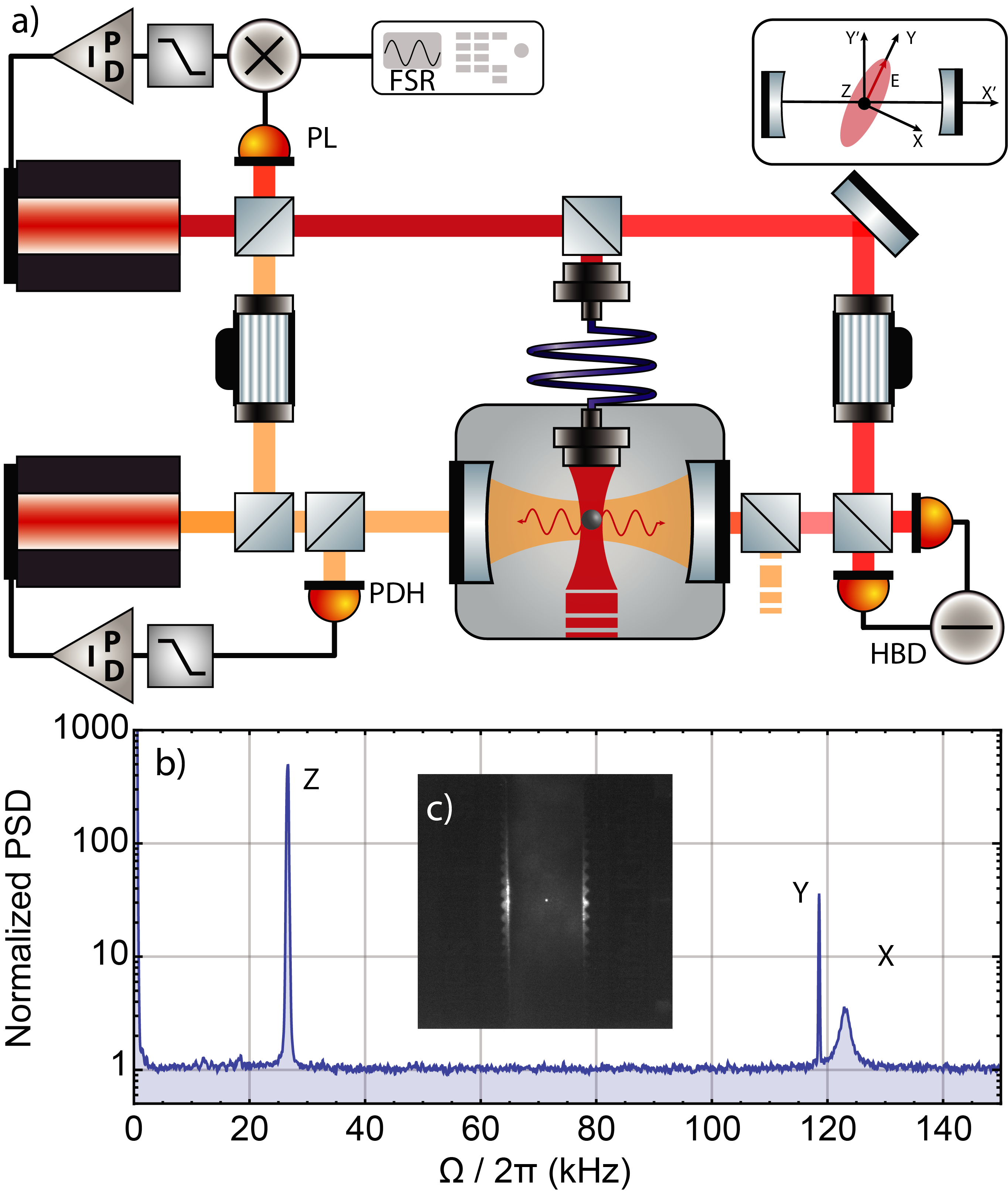}
\caption{a) Simplified scheme of the experimental setup. PL: Phase locking, PDH: Pound-Drever-Hall detection, HBD: heterodyne balanced detection. (b) Spectrum of the three-dimensional motion of the nanosphere when it is located close to the node of the standing wave inside the optical cavity, at the background pressure of $3 \times 10^{-5}$ Pa. The spectrum is obtained from the heterodyne detection of detuned tweezer light (-$\Delta/2\pi = 260$ kHz), scattered by the nanosphere and transmitted by the cavity, and it is normalized to the detection shot-noise. The peaks corresponding to the oscillations along the three orthogonal axes ($Z$ along the tweezer propagation, $Y$ along its polarization axis, and $X$) are visible around respectively 27, 118 and 123 kHz. (c) Photo of the levitating nanosphere.}
\label{fig_setup}
\end{figure} 

The nanosphere is positioned inside a nearly-concentric Fabry-Perot cavity (free-spectral-range $FSR = 3.07$ GHz, linewidth $\kappa/2\pi = 57$ kHz), orthogonal to the tweezer axis (Fig. \ref{fig_setup}). A second, auxiliary Nd:YAG laser is frequency locked to the optical cavity, and the tweezer laser is phase locked to the auxiliary one with a controllable frequency offset equal to $FSR + \Delta/2\pi$, thus accurately defining the detuning $\Delta$ of the tweezer radiation from the cavity resonance. 
The light scattered by the particle on the cavity mode and transmitted by the output mirror is superimposed on a local oscillator beam, derived from the main Nd:YAG laser before launching it in the fiber, and frequency shifted by 1.1 MHz. The mixed beams are sent to a balanced detection to implement a heterodyne measurement. The beat note between the local oscillator and the scattered light allows to deduce the location of the nanosphere inside the mode standing wave, and to place it on the optical axes, in correspondence of a node. In this position, the optical coupling between the particle motion and the radiation field is due to the coherent scattering on the cavity mode and it is effective just for the projection of the motion on the cavity axis. The optomechanical coupling rates for the oscillations parallel ($Y$ direction) and perpendicular ($X$ direction) to the tweezer polarization are $\gX\,=\,g\,\sin^2 \theta$ and $\gY\,=\,g\,\cos \theta \sin \theta$ where $\theta$ is the angle between the cavity and the polarization axes. Our measured values are $\theta = 72^{\circ}$ and $g = 2\pi \times 30$ kHz. The gas damping rate $\Gm$ is proportional to the pressure $P$ (for high Knudsen number, i.e., below $\sim 1$ kPa) and for our conditions (170 nm diameter silica nanospheres in pure nitrogen atmosphere) is about $\Gm/2\pi \simeq 10$ Hz $\times P$(Pa). The coherent strong coupling regime is expected to be achieved when $\,2g > \Gamma_{\mathrm{th}} = \Gamma_{\mathrm{m}} n_{\mathrm{th}} + \Gamma_{\mathrm{n}}$, where $n_{\mathrm{th}}$ is the mean occupation number of the thermal bath, and $\Gamma_{\mathrm{n}}$ is the decoherence rate additional to the gas damping.

\begin{figure}
\includegraphics[width=0.95\columnwidth]{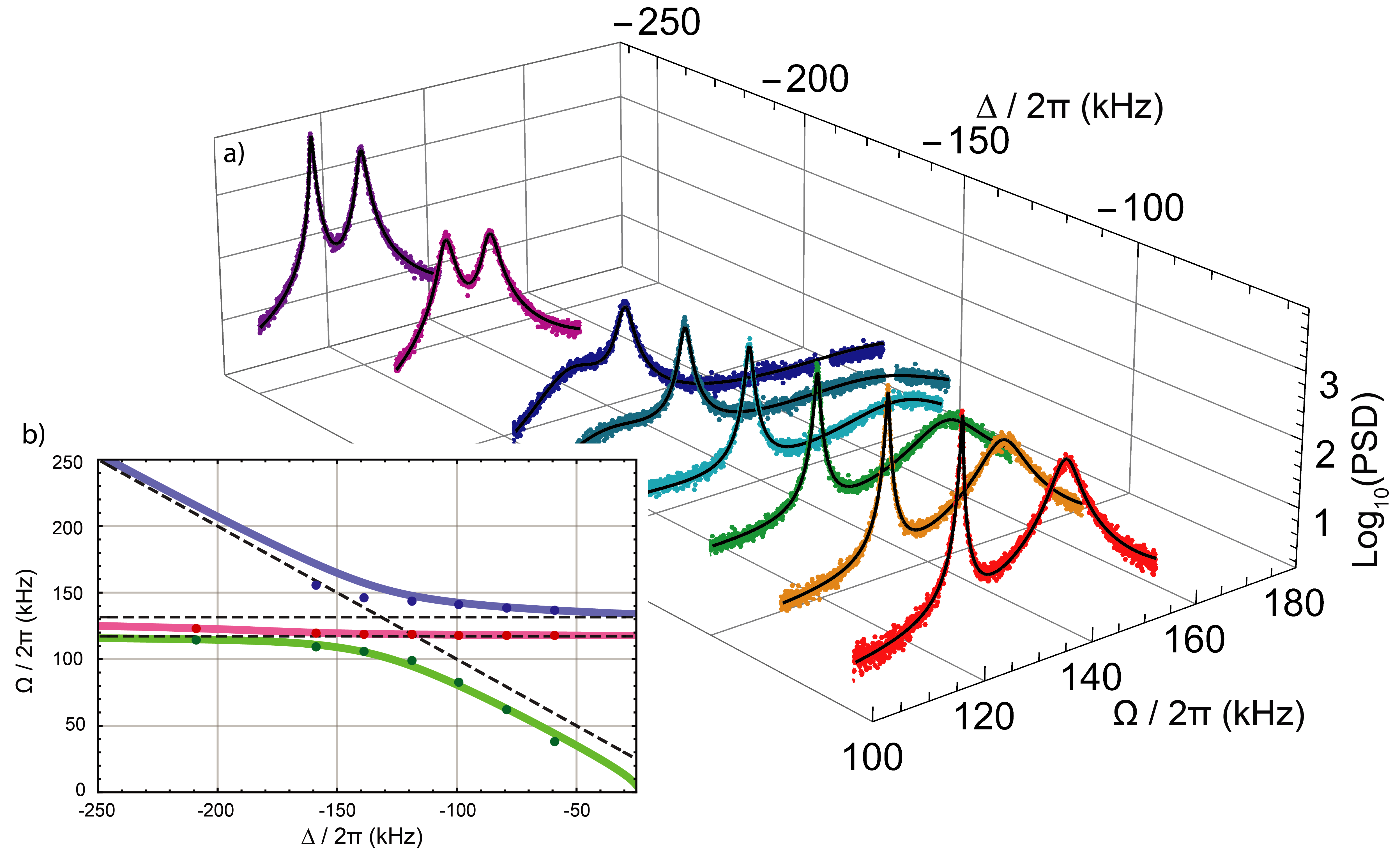}
\caption{a) High frequency sideband of the heterodyne spectra, for a background pressure of $\sim 6 \times 10^{-3}$ Pa. The curves are shown for decreasing detuning of the tweezer light from the cavity resonance, namely for values of $-\Delta/2\pi = (260, 210, 160, 140, 120, 100, 80, 60)$ kHz. Black solid lines show the theoretical spectra. (b) Dispersion curves obtained, as a function of $\Delta$, from the imaginary part of the eigenvalues of the drift matrix (solid lines). The experimental heterodyne spectra do not provide a direct access to such eigenvalues, however their shape can be approximated as a sum of three Lorentzian functions, whose centers, obtained from a fit, are shown with dots in the figure.}
\label{fig_spettri}
\end{figure}

In Fig. \ref{fig_spettri}a we show the spectra (high frequency sideband) of the transmitted field when the pressure in the experimental chamber is about $6 \times 10^{-3}$ Pa, for varying detuning $\Delta$ of the tweezer light with respect to the cavity resonance. The photonic component becomes evident for  $-\Delta/2\pi = 160$ kHz, where the spectrum shows three peaks for the optical and the two mechanical resonances. The $X$ and $Y$ mode peaks merge, forming the dark and bright modes and, at slightly smaller detuning, the bright mode strongly hybridizes with the photon field, forming the polaritons. At $-\Delta/2\pi \simeq 140$ kHz we clearly distinguish the photon-like polariton (on the high frequency side of the dark mode peak), and the phonon-like polariton (on the opposite side). For $-\Delta/2\pi \simeq 120$ kHz both polaritons are almost equally composed of phononic and photonic components and, as we will see, the minimum occupation number is achieved. 
The corresponding resonances have width $\sim \kappa/2$ and are separated by $\sim 2 g$. 
Between them we see the narrow resonance originated by the dark mode. For smaller detuning, the relative position of the photon-like and phonon-like polariton peaks are swapped. 

The dispersion relation is reported in Fig. \ref{fig_spettri}b.  The polaritonic branches are separated by an energy gap, originated by an upper and a lower avoided-crossing, and asymptotically approach the correspondent free mechanical frequencies. This situation is strongly reminiscent of phonon-polaritons in ionic crystals, where the asymptotic frequencies are those of the longitudinal and transverse optical phonons \cite{huang}. Inside the gap the dispersion curve of the dark mode takes place, similarly to what is observed in the spectra of two-qubit states interacting with microwave photons \cite{sun,altomare}.

The Hamiltonian of the three interacting oscillators (one optical and two mechanical) is 
\begin{equation}
H = -\hbar \Delta a^{\dagger} a + \hbar\Omega_1  b^{\dagger}_1 b_1+ \hbar \Omega_2  b^{\dagger}_2 b_2+ \hbar g_1 (a^{\dagger} +a)(b^{\dagger}_1+b_1) + \hbar g_2 (a^{\dagger}+a)(b^{\dagger}_2+b_2) \, .
\label{hamiltonian}
\end{equation}
where $a$ ($b_{i}$, $i=1, 2$) are the bosonic operators of the optical (mechanical) oscillators, and $\Omega_{i}$ the mechanical angular frequencies.
A quantum Langevin model is derived from the Hamiltonian by adding the input terms. In particular, for the mechanical degrees of freedom we take into account the coupling with the thermal bath by means of gas damping, and the heating due to the shot noise in the dipole scattering, that is relevant in high vacuum. From the Langevin model and the input/output relation for the transmitted field we calculate the heterodyne spectra, that are shown in Fig. \ref{fig_spettri} and display an excellent agreement with the experimental data. The frequency and optomechanical gains are adapted to the experimental data, while all the other parameters are measured independently. Even if a direct measurement of the nanosphere motion is not possible, since we just have access to the strongly interacting optical meter, the agreement between theory and experiment justifies the assumption that the system is well modelled and consequently the trust in the theoretical calculations. The phonon occupation numbers for the $X$ and $Y$ mechanical modes are calculated from the integral of the respective spectra. The minimal values are achieved for a detuning of 120 kHz and are respectively around 200 and 1500. The coldest oscillation is not however along the $X$ direction, but at about $15^{\circ}$, close to the cavity axis, and its effective occupation number is as low as 100. 

\begin{figure}
\includegraphics[width=0.8\columnwidth]{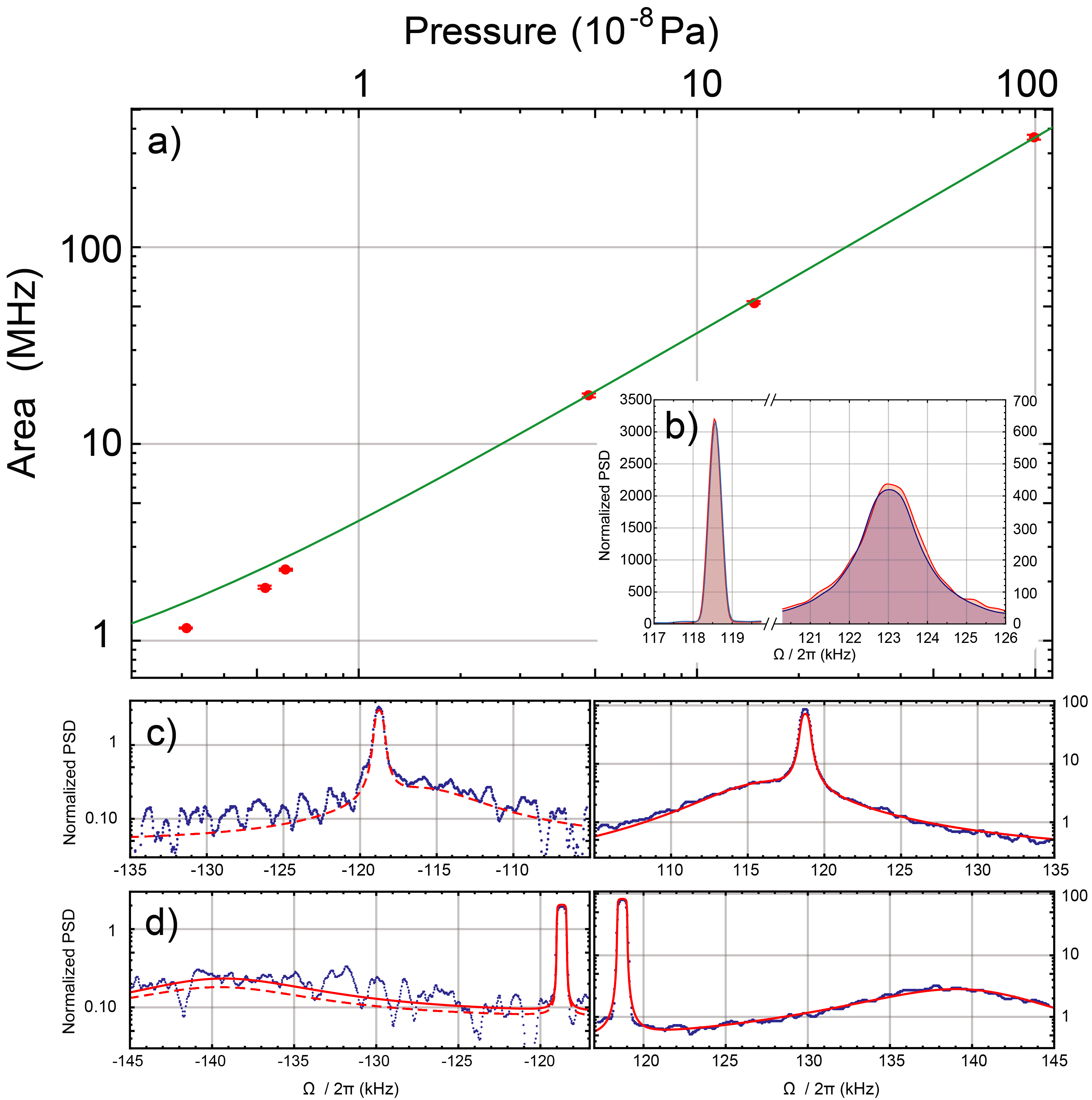}
\caption{a) Area of the spectral peak corresponding to the motion along the $X$ direction (high frequency sideband in the heterodyne detection), measured for a tweezer light detuning of $-\Delta/2\pi = 260$ kHz as a function of the background pressure. The spectra are normalized to the detection shot noise. The solid line shows the fitted linear interpolation. (b) Spectra of the two motional sidebands, multiplied by the cavity filtering function $1+\left(\frac{\Delta+\Omega}{\kappa/2}\right)^2$, acquired at $3 \times 10^{-5}$ Pa. The hot $Y$ mode generate symmetric peaks around 118.7 kHz, while the broad peaks corresponding to the $X$ mode show an asymmetry in agreement with the phononic occupation number of $\sim 20$, calculated by taking into account gas damping and dipole light scattering. (c-d) Enlarged views of the heterodyne spectra for a detuning of $-\Delta/2\pi = 170$ kHz (c) and $-\Delta/2\pi = 100$ kHz (d), showing the narrow dark mode and the broad phonon-like polariton. The zero frequency is set at the difference between tweezer light and local oscillator (i.e., 1.1 MHz). The spectra are normalized to shot noise, that was then subtracted, and a moving average with a width of 600 Hz is applied. Solid lines show the model prediction. On the low-frequency sideband, the dashed line shows the prediction in case of classical, symmetric noise spectrum, i.e., the high-frequency sideband multiplied by $\frac{(\kappa/2)^2+(\Delta-\Omega)^2}{(\kappa/2)^2+(\Delta+\Omega)^2}$.} 
\label{fig_area}
\end{figure}

The effective temperature of the particle motion is further reduced by decreasing the background pressure and consequently the gas damping. This is well characterized by the spectra at large detuning (we set at this purpose $-\Delta/2\pi = 260$ kHz), where the peak of the $X$ mode remains clearly distinguishable. Its area decreases linearly with pressure in the mid-vacuum range (see Fig. \ref{fig_area}a), and from the slope we deduce $\Gm/2\pi P = 13$ Hz/Pa in agreement with the expectation. In high vacuum, pressure-independent decoherence terms become relevant, in particular the shot-noise in the dipole scattering \cite{Jain:2016} that is calculated to be $\Gamma_{\mathrm{sc}}/2\pi = 8.9$ kHz for the $X$ mode and 4.8 kHz for the $Y$ mode \cite{Seberson:2019}. The experimental data extend down to $3 \times 10^{-5}$ Pa without showing effects of possible technical noise (such as parametric heating or mechanical vibrations \cite{gonzalez-ballestrero2019}), as also testified by the slight asymmetry in the motional sidebands (Fig. \ref{fig_area}b). At this point, the optomechanical gain $2\gX = 2\pi \times 54$ kHz exceeds the total decoherence rate $n_{\mathrm{th}} \Gm + \Gamma_{\mathrm{sc}} = 2 \pi \times 38$ kHz more than in any system based on optical fields, in spite of the room-temperature operation.  

By decreasing the detuning, the system fully enters the quantum coherent strong-coupling regime. Around $-\Delta/2\pi \simeq 130$ kHz, where the phonon-like and photon-like polariton have similar strength in the output field, their spectral feature is too flat and close to the detection shot-noise level to provide a reliable quantitative information. As a consequence, we focus here on two values of detuning, namely $-\Delta/2\pi = 170$ kHz and 100 kHz, where the peak of the phonon-like polariton is clearly visible respectively at lower and higher frequency with respect to that of the dark mode (Fig. \ref{fig_area}c-d). The asymmetry in the spectra of the motional sidebands is here a signature of the non-classicality. The phonon-like polariton is 78 (76) $\%$ phononic and 22 (24) $\%$ photonic at $-\Delta/2\pi = 170$ kHz ($100$ kHz). In the coldest direction, the inferred occupation number is respectively 5 and 2.5, and it decreases to $\sim 1.5$ for $-\Delta/2\pi \simeq 120$ kHz. 

We finally comment on the perspectives opened by our results. The preparation of quantum polaritons here demonstrated is the pre-requisite for transferring quantum information between photonic and phononic components. The dark mode, weakly interacting with both the photonic field and (at low pressure) the thermal bath, is suitable for its long-term storage. The dispersion relation displays two avoided-crossings, as typically observed in tripartite quantum systems. Notably, each avoided crossing acts as a quantum beam splitter of wave functions, driving an input quantum state into a coherent superposition of two output states evolving independently in time \cite{sun,petta}. Beam-splitters are basic components to realize a number of quantum operations, such as entanglement \cite{tan,kim} and teleportation \cite{wang}. The realized system thus paves the way to novel protocols for the quantum coherent control of phononic and photonic modes and represents a key-step towards the demonstration optomechanical entangled states at room temperature. Furthermore, the phonon-polaritons form an useful basis for developing non-linear quantum optomechanics \cite{liu2013,lemonde2013,borkje2013}.

\section*{Acknowledgements}
F.M. and P.V. would like to thank N. Kiesel, U. Delic and M. Aspelmeyer for useful discussions and their kind hospitality.
Research performed within the Project QuaSeRT funded by the QuantERA ERA-NET
Cofund in Quantum Technologies implemented within the European Union's
Horizon 2020 Programme.

\section*{Appendix}

\subsection*{Trapping of the nanosphere}

We load the silica nanosphere (170 nm diameter) in a first vacuum chamber, on a first optical tweezer mounted on the tip of a movable rod \cite{mestres2015,calamai2020}. The trapped particle is then moved to the experimental chamber, where it is transferred to a second tweezer, formed by the light of a Nd:YAG laser delivered in vacuum by an optical fiber \cite{calamai2020}. The photo in Fig. \ref{fig_setup}c shows the nanosphere during the transfer between the two tweezers. About 300 mW are focused to an elliptical shape of waists $1.02 \mu$m and $0.93 \mu$m, where the tighter focus occurs in the direction orthogonal to the axis defined by the linear polarization of the tweezer light \cite{Novotny:2012}. The corresponding typical oscillation frequencies of the nanosphere in the tweezer optical potential are  27 kHz ($Z$ direction, along the tweezer axis), 120 kHz ($Y$ direction, along the polarization axis) and 130 kHz ($X$ direction). The second tweezer is mounted on a three-axes nanometric motorized linear stage.  After the transfer, the movable rod is retracted, the two chambers are isolated and the experimental chamber is pumped to high vacuum.

\subsection*{Theoretical model}

The standard quantum Langevin model describing the optomechanical interaction between a mechanical modes and the cavity field can be easily extended to the two-dimensional motion of the nanosphere as follows
\begin{equation}
\dot{\hat{a}}_{\mathrm{c}}=\bigg(  i\Delta-\frac{\kappa}{2}\bigg)\ac+i\gX(\hat{b}_X+\hat{b}_X^\dag)+i\gY(\hat{b}_Y+\hat{b}_Y^\dag)+\sqrt{\kappa}\,\hat{a}_{\mathrm{in}}
\label{dta}
\end{equation}
\begin{equation}
\dot{\hat{b}}_j=\left(-i\Omegamj^0-\frac{\Gmj}{2}\right)\hat{b}_j+ig_j(\ac+\ac^{\dag})+\sqrt{\Gmj}\,\hat{b}_{\mathrm{th},j}+\sqrt{\Gnj}\,\hat{b}_{\mathrm{n},j}
\label{dtb}
\end{equation}
where the linearized evolution equations are expressed in the frame rotating at the laser frequency $\omega_{\mathrm{L}}$, the bosonic field operators $\ac$ and $\hat{b}_j$ describe respectively the intracavity field and the two mechanical modes ($j=X, Y$),  $\Delta=\omega_{\mathrm{L}}-\omega_{\mathrm{c}}$ is the detuning with respect to the cavity resonance frequency $\omega_{\mathrm{c}}$, $\kappa$ and $\Gmj$ are the optical and mechanical decay rates, $\Omegamj^0$ are the mechanical resonance frequencies, and $g_j$ the opto-mechanical coupling rates.
The input noise operators are characterized by the correlation functions
\begin{eqnarray}
\langle\hat{a}_{\mathrm{in}}(t)\hat{a}_{\mathrm{in}}^{\dag}(t')\rangle & = & \delta(t-t')
\label{noise1} \\
\langle\hat{a}_{\mathrm{in}}^{\dag}(t)\hat{a}_{\mathrm{in}}(t')\rangle & = & 0
\label{noise2}  \\
\langle\hat{b}_{\mathrm{th},j}(t)\hat{b}_{\mathrm{th},j}^{\dag}(t')\rangle & = & (\bar{n}_{\mathrm{th},j}+1)\,\delta(t-t')
\label{noise3}    \\
\langle\hat{b}_{\mathrm{th},j}^{\dag}(t)\hat{b}_{\mathrm{th},j}(t')\rangle & = & \bar{n}_{\mathrm{th},j}\,\delta(t-t')
\label{noise4}     \\
\langle\hat{b}_{\mathrm{n},j}^{\dag}(t)\hat{b}_{\mathrm{n},j}(t')\rangle = \langle\hat{b}_{\mathrm{n},j}(t)\hat{b}_{\mathrm{n},j}^{\dag}(t')\rangle & = & \delta(t-t')
\label{noise5}
\end{eqnarray}
where $\bar{n}_{\mathrm{th},j} = \left(\exp\left(\frac{-\hbar \Omegamj^0}{\kB T}\right)-1\right)^{-1}$ are the thermal occupation numbers of the two mechanical modes and $b_{\mathrm{n},j}$ accounts for additional heating sources, such as dipole scattering, parametric heating, trap vibrations.  

In the parameters range that assures the system stability, the stationary evolution equations can be written in the Fourier space in the compact matrix form as
\begin{equation}
\left(-i \Omega \,\mathbf{I} + \mathbf{D}\right) \mathbf{V} = \mathbf{V}_{\mathrm{in}} 
\label{eq_Fourier}
\end{equation}
where $\mathbf{I}$ is the identity matrix of order 6, the drift matrix is
\begin{equation*}
\mathbf{D}=
\begin{scriptscriptstyle}
\left(\begin{array}{cccccc}
-i \Delta+\kappa/2 & 0 & -i \gX & -i \gX & -i \gY & -i \gY \\
0 & i \Delta+\kappa/2 &i \gX & i \gX & i \gY & i \gY \\
-i \gX & -i \gX & i \OmegaX+\GmX/2 & 0 & 0 & 0 \\
i \gX & i \gX & 0 & -i \OmegaX+\GmX/2 & 0 & 0  \\
-i \gY & -i \gY & 0 & 0 & i \OmegaY+\GmY/2 & 0 \\
i \gY & i \gY & 0 & 0 & 0 & -i \OmegaY+\GmY/2  \\
\end{array} \right) 
\end{scriptscriptstyle}
\, ,
\end{equation*}
the intracavity vector is $\mathbf{V} = (\tac,\tac^{\dagger},\tbX,\tbdX,\tbY,\tbdY)$ where $\tilde{O}(\Omega)$ is the Fourier transformed of the operator $\hat{O}(t)$ and $\tilde{O}^{\dagger}(\Omega) = \left(\tilde{O}(-\Omega)\right)^{\dagger}$ is the Fourier transformed of $\hat{O}^{\dagger}(t)$. The input noise vector $\mathbf{V}_{\mathrm{in}}$ is defined as $\mathbf{V}_{\mathrm{in}}=(\sqrt{\kappa}\, \tilde{a}_{\mathrm{in}},\sqrt{\kappa}\, \tilde{a}_{\mathrm{in}}^{\dag},\sqrt{\GmX} \,\tilde{b}_{\mathrm{th},X}+\sqrt{\GnX} \,\tilde{b}_{\mathrm{n},X},\sqrt{\GmX} \,\tilde{b}_{\mathrm{th},X}^{\dag}+\sqrt{\GnX} \,\tilde{b}_{\mathrm{n},X}^{\dag},\sqrt{\GmY} \,\tilde{b}_{\mathrm{th},Y}+\sqrt{\GnY} \,\tilde{b}_{\mathrm{n},Y},\sqrt{\GmY} \,\tilde{b}_{\mathrm{th},Y}^{\dag}+\sqrt{\GnY} \,\tilde{b}_{\mathrm{n},Y}^{\dag})$, and $\mathbf{V}$ is found by inverting Eq. (\ref{eq_Fourier}), i.e., $\mathbf{V} = \left(-i \Omega \,\mathbf{I} + \mathbf{D}\right)^{-1} \mathbf{V}_{\mathrm{in}}$.  

If we want to describe the nanosphere motion using a reference frame rotated by an angle $\theta$ with respect to the initial one, we can apply the rotation matrix to the coordinates ($\bX+\bdX$) and ($\bY+\bdY$). 

The total cavity output field is $\hat{a} = \sqrt{\kappa} \hat{a}_{\mathrm{c}}-\hat{a}_{\mathrm{in}}$, and we can define an output vector $\mathbf{V}_{\mathrm{out}}=(\tilde{a},\tilde{a}^{\dagger},\tbX,\tbdX,\tbY,\tbdY)$ that is linked to the input noise vector by the equation $\mathbf{V}_{\mathrm{out}} =  \mathbf{O}\,\mathbf{V}_{\mathrm{in}}$ where the final output matrix is
\begin{equation*}
\mathbf{O} =  
\left(\begin{array}{cccccc}
\sqrt{\kappa} & 0 & 0 & 0 & 0 & 0 \\
0 & \sqrt{\kappa} & 0 & 0 & 0 & 0 \\
0 & 0 & \cos{\theta} & 0 & \sin{\theta} & 0  \\
0 & 0 & 0 & \cos{\theta} & 0 &  \sin{\theta} \\
0 & 0 & - \sin{\theta} & 0 & \cos{\theta} & 0   \\
0 & 0 & 0 & - \sin{\theta} & 0 & \cos{\theta} \\
\end{array} \right) 
\left(-i \Omega \,\mathbf{I} + \mathbf{D}\right)^{-1}\,-\,
\left(\begin{array}{cccccc}
1/\sqrt{\kappa} & 0 & 0 & 0 & 0 & 0 \\
0 & 1/\sqrt{\kappa} & 0 & 0 & 0 & 0 \\
0 & 0 & 0 & 0 & 0 & 0  \\
0 & 0 & 0 & 0 & 0 & 0   \\
0 & 0 & 0 & 0 & 0 & 0   \\
0 & 0 & 0 & 0 & 0 & 0 \\
\end{array} \right)
\, .
\end{equation*}
The output spectra are calculated by taking just the coefficients of the few non-null terms in the input noise correlation functions. By calling $O^{\pm} [i,j]$ the (i,j) element of $\mathbf{O} (\pm\Omega)$, a generic output spectrum can be written in the form
\begin{eqnarray*}
S_{i,j} &=& \kappa \,O^-[i,1] O^+[j,2] \\
& + & \left(\GmX(\bar{n}_{\mathrm{th},X}+1)+\GnX \right) O^-[i,3] O^+[j,4] 
 +  \left(\GmX \bar{n}_{\mathrm{th},X} + \GnX \right)O^-[i,4] O^+[j,3]  \\
& + & \left(\GmY(\bar{n}_{\mathrm{th},Y}+1)+\GnY \right) O^-[i,5] O^+[j,6]+  \left(\GmY \bar{n}_{\mathrm{th},Y} +\GnY \right) O^-[i,6] O^+[j,5]         \, .       
\end{eqnarray*}

For the output field we obtain
\begin{eqnarray*}
S_{a a} &=& \frac{1}{2\pi}\langle \tilde{a}^{\dagger}(-\Omega) \tilde{a}(\Omega)\rangle \,\equiv\, S_{2,1}       \\
S_{a^{\dagger} a^{\dagger}} &\equiv& S_{1,2}
\end{eqnarray*}
and the heterodyne spectrum, normalized to shot noise, is
\begin{equation}
S_{\mathrm{out}} = \Big(S_{a a}(\Omega-\Omega_{\mathrm{LO}})+S_{a^{\dagger} a^{\dagger}}(\Omega+\Omega_{\mathrm{LO}})\Big) \eta+ (1-\eta)
\label{eq_Sout}
\end{equation}
where $\Omega_{\mathrm{LO}}$ is the angular frequency of the local oscillator, and $\eta$ is the detection efficiency.

The spectrum of the oscillations along the direction defined by $\theta$ is obtained as $S_{b b}+S_{b^{\dagger} b^{\dagger}}$ where
\begin{eqnarray}          
S_{b b} &\equiv& S_{4,3}      \label{eq_Sbb} \\
S_{b^{\dagger} b^{\dagger}} &\equiv& S_{3,4}
\end{eqnarray}
and it can be shown that, since the bosonic commutation relations are conserved by the axes rotation, $\int S_{b^{\dagger} b^{\dag}} \,\frac{\ud \Omega}{2 \pi} = 1+ \int S_{b b} \,\frac{\ud \Omega}{2 \pi}$.
The spectra of the motion along the original $X$ and $Y$ axes are obviously recovered by setting respectively $\theta=0$ and $\theta=\pi/2$, while the angle giving the coldest motion, i.e., the minimum displacement spectrum, is obtained by the relation
\begin{equation}
\theta_{\mathrm{min}}\,=\,\frac{1}{2} \arctan \frac{\int S_{\mathrm{XY}}\,\frac{\ud \Omega}{2 \pi}}{\int \left(S_{b b}(\theta\rightarrow\pi/2)-S_{b b}(\theta\rightarrow 0)\right)\,\frac{\ud \Omega}{2 \pi}}
\end{equation}
where the cross-correlation spectrum is 
$S_{\mathrm{XY}} \,\equiv\, S_{6,3}$  .

Equation (\ref{eq_Sout}) is used to derive all the theoretical spectra shown in this article, while Eq. (\ref{eq_Sbb}) allows to calculate the occupation number for the oscillations along the direction defined by $\theta$, according to $\nm = \int S_{b b} \,\frac{\ud \Omega}{2 \pi}$.

From Eq. (\ref{dta}) we derive that the output field can be written in the form 
\begin{equation}
\tilde{a}\,=\,\left(\frac{\kappa}{-i\Omega-i\Delta+\kappa/2}-1\right)\tilde{a}_{\mathrm{in}}\,+\,\frac{i\gX(\tilde{b}_X+\tilde{b}_X^\dag)+i\gY(\tilde{b}_Y+\tilde{b}_Y^\dag)}{-i\Omega-i\Delta+\kappa/2}
\end{equation}
where the second term on the right hand side is suggestive of the spectrum of the oscillators filtered by the cavity, that could therefore be directly inferred from the field spectrum. However, we remark that this simplified view, while catching the main features of the output spectrum, fails to fully consider the correlation between the motion and $\hat{a}_{\mathrm{in}}$, that becomes particularly important in case of strong coupling. The full model is necessary to correctly derive the output spectra, and the behavior of single components of the system (such as the nanoparticle motion) can just be extracted from the model, provided that it accurately describe the experimental observations.

\begin{figure}
\includegraphics[width=0.98\columnwidth]{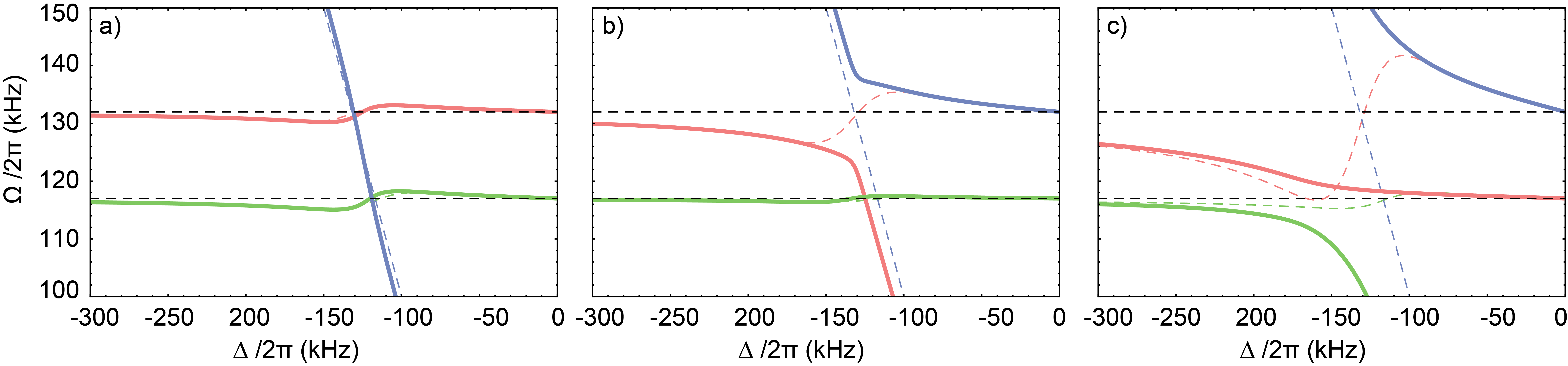}
\caption{Solid lines: eigenfrequencies derived from the imaginary part of the eigenvalues of the drift matrix, as a function of the detuning. Colored dashed lines: effective frequencies calculated in the weak coupling approximation (Eq. \ref{eq_Weff}). Black dashed lines: natural mechanical frequencies. Common parameters are $\kappa/2\pi = 57$ kHz, $\Gm/2\pi = 0.1$ Hz, $\OmegaX/2\pi = 132$ kHz, $\OmegaY/2\pi = 117$ kHz. The optomechanical gains are $\gX = \gY = 2 \pi \times 9$ kHz (a), $\gX = 2 \pi \times 16$ kHz; $\gY = 2 \pi \times 5$ kHz (b), $\gX = 2 \pi \times 29$ kHz; $\gY = 2 \pi \times 9$ kHz (c).}
\label{fig_teo}
\end{figure}

The essential information on the system dynamical behavior is encoded in the eigenvalues and eigenvectors of the drift matrix. The formers are three couples of complex conjugate parameters, where the absolute values of the imaginary parts give the three eigenfrequencies, and the corresponding real parts gives the widths of the resonances. For weak coupling ($\gX, \gY \ll \kappa/2$) and well separated mechanical frequencies ($|\OmegaX - \OmegaY| \gg 4 g_{\mathrm{X,Y}}^2/\kappa$) the optical eigenvalues are $\sim \kappa/2 \pm i \Delta$ and the mechanical resonances are given approximately by \cite{Aspelmeyer:2014_review}
\begin{equation}
\Omega_{\rm{eff}}\,\simeq \Omega^0\,+\,g^2 \left( \frac{\Delta-\Omega^0}{(\Delta-\Omega^0)^2 + \kappa^2/4} + \frac{\Delta+\Omega^0}{(\Delta+\Omega^0)^2 + \kappa^2/4}\right)
\label{eq_Weff}
\end{equation}
\begin{equation}
\Gamma_{\rm{eff}}\,\simeq \Gamma_{\rm{m}}\,+\,\,g^2 \left( \frac{\kappa}{(\Delta+\Omega^0)^2 + \kappa^2/4} - \frac{\kappa}{(\Delta-\Omega^0)^2 + \kappa^2/4}\right)
\label{eq_Geff}
\end{equation}
where we have omitted the subscripts X,Y (Fig. \ref{fig_teo}a).

When the optical spring effect is strong enough, and the two natural mechanical frequencies close enough, that the effective frequencies of the two oscillators get closer than their effective widths, the two mechanical modes hybridize, and two linear combinations exhibit the highest and lowest coupling to the optical field. They are, respectively, the bright and dark mode: $b_{\mathrm{bright}} \propto \gX \bX + \gY \bY$ and $b_{\mathrm{dark}} \propto \gX \bX - \gY \bY$ \cite{genes2008,massel2012,shkarin}. Such modes are not exact eigenvectors of the systems, therefore the dark mode resonance is broadened by the coupling with the bright mode (and actually with the optical field), however a narrow resonance remains the signature of the hybridization. A drawback of the hybridization is that the optical cooling of the bright mode is hindered by its coupling to the hot, dark mode, therefore reaching the quantum regime is more difficult \cite{genes2008}. On the other hand, hybridization allows the transfer of energy and information between the two mechanical modes.     

If the two mechanical frequencies are well separated, when one mode has a coupling strength $g > \kappa/4$ it enters the strong optomechanical coupling regime, characterized by an avoided crossing between the eigenfrequencies of the optical  mode (that significantly departs from $\Delta$) and of the mechanical mode (Fig. (\ref{fig_teo}b). The width of the latter does not increase as much as predicted by Eq. (\ref{eq_Geff}), and it is upper limited to $\kappa/2$. Like the mechanical hybridization, also this effect hinders the attainment of the mechanical quantum regime, even if a phononic occupation number below unity has been achieved even in the strong coupling regime in ultra-cryogenic experiments with a microwave coupled to an aluminum drum oscillator \cite{wollman,pirkka}. 

Finally, in the strong coupling regime involving both mechanical modes, full development of the system into vectorial polaritons occurs, as described in the main text (Fig. (\ref{fig_teo}c). The threshold to reach strong coupling (and avoided crossing) is lowered, approaching $\gX \gY > \kappa^2/32$ for $\OmegaX \simeq \OmegaY$.

\subsection*{Experimental setup}

\begin{figure}
\includegraphics[width=0.8\columnwidth]{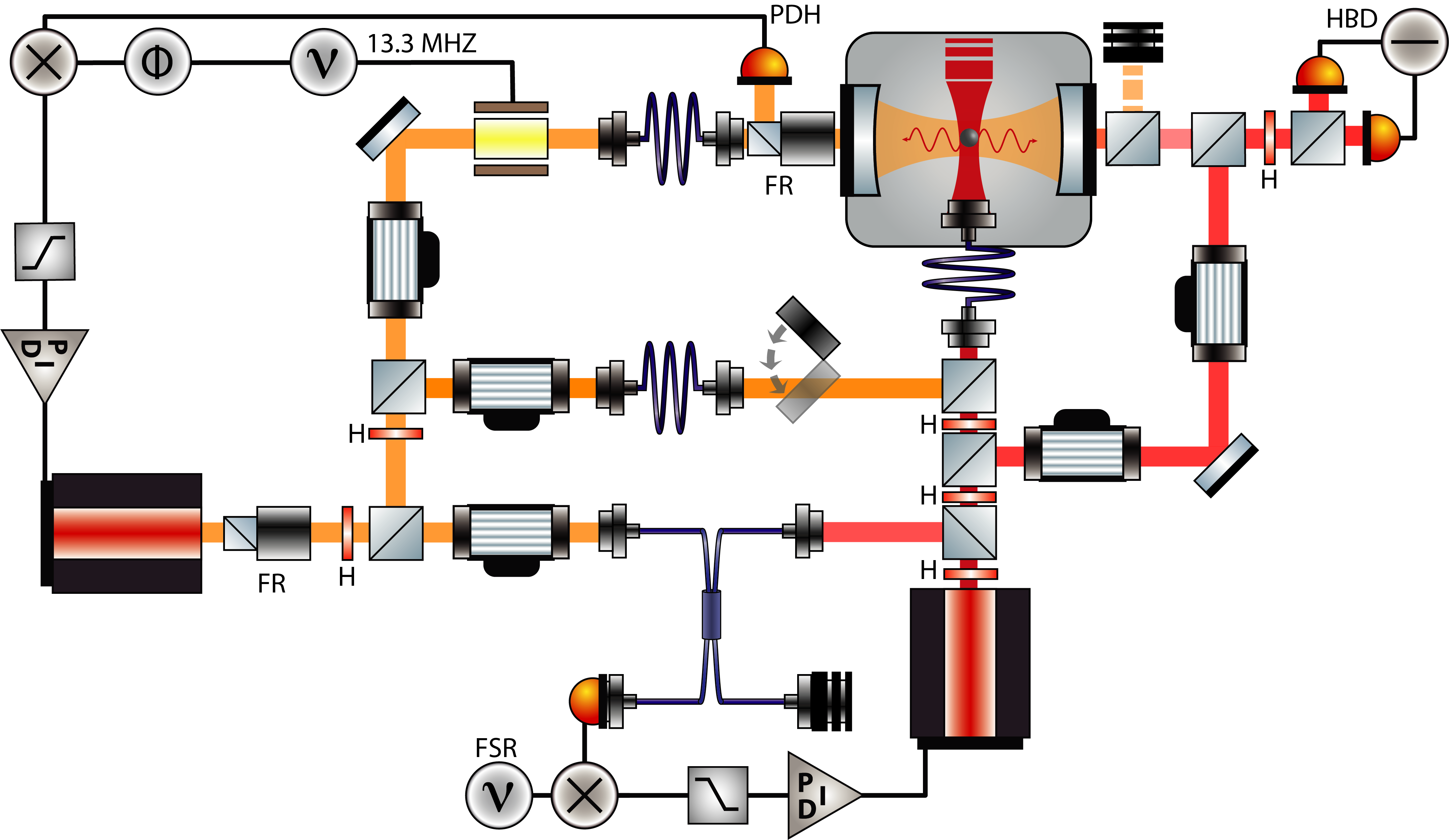}
\caption{Scheme of the experimental setup. PDH: Pound-Drever-Hall detection, HBD: heterodyne balanced detection, FR: Faraday rotator, H: half-way plate, FSR: sinusoidal wave generator at frequency corresponding to $\sim 1$ cavity free-spectral-range.}
\label{fig_setup2}
\end{figure}

A simplified scheme of the experimental setup is shown in Fig. (\ref{fig_setup2}). From a first Nd:YAG laser we derive three beams, all frequency shifted by acousto-optic modulators and sent to the experimental bench by polarization-maintaining optical fibers. The first beam (reference) is phase modulated at 13.3 MHz, mode-matched to the cavity and used to frequency lock the laser to a cavity resonance by means of the Pound-Drever-Hall technique. The second beam is used for phase locking the second Nd:YAG laser (Nd:YAG2). At this purpose, it is mixed in a fibered beam-splitter with the radiation derived from Nd:YAG2, and detected by a fast photodiode. The beat note is down-converted in a mixer by a local oscillator at $\sim3$ GHz (one cavity free-spectral-range), and the IF output of the mixer is used in a servo loop acting on Nd:YAG2 for phase locking. The third beam is superposed to the main tweezer light (supplied by Nd:YAG2), red detuned from a cavity resonance and used for stabilization purposes during the pumping down. The presence of two fields in the tweezer, red detuned from consecutive cavity resonances, provides indeed cooling along all the three directions for any position of the nanosphere, thanks to the two shifted standing waves. This third beam is blocked during the measurements.  

Two weak beams are derived from Nd:YAG2. The first one is launched into an optical fiber and used for phase locking, as just described, the second one is frequency shifted by 1.1 MHz using a cascade of two acousto-optic modulators working on opposite orders, and it provides the local oscillator in a balanced heterodyne detection. The light transmitted by the cavity is mode-matched and superposed to the local oscillator with orthogonal polarization, in a first polarizing beam-splitter. The resulting beam is then directed to the balanced detection setup, composed of a half-wave plate, a second polarizing beamsplitter, and a couple of photodetectors, whose signals are electronically subtracted. 

\begin{figure}
\includegraphics[width=0.8\columnwidth]{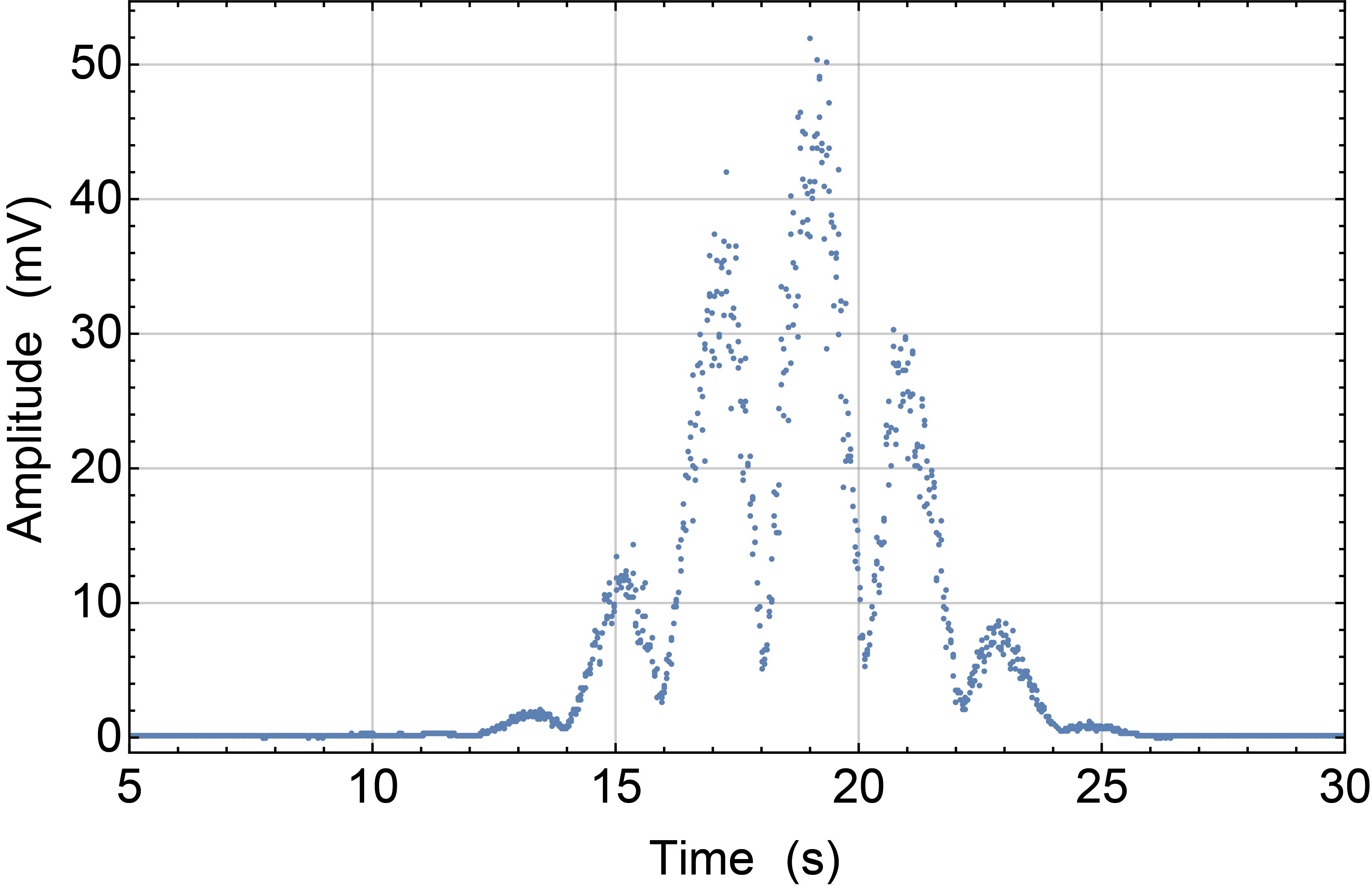}
\caption{Amplitude of the beat note at 1.1 MHz between the tweezer light, scattered by the nanosphere and transmitted through the cavity output mirror, and the local oscillator. The nanosphere is moved through the cavity standing wave using the tweezer positioner.}
\label{fig_battimento}
\end{figure}

Most of the Nd:YAG2 power is launched into a polarization maintaining fiber and used for the optical tweezer. The fiber is delivered into the experimental vacuum chamber, where its output is collimated and re-focused by a doublet of aspheric lenses (respectively with focal length 15.4 mm, NA 0.16 and focal length 3.1 mm, NA 0.68), mounted on the fiber FC connector. The fiber tip with the optical system is movable along three axes using nanometric vacuum positioners. Its displacement is calibrated by observing the beat note between the light scattered by the nanosphere on the cavity mode and transmitted through the output mirror, and the local oscillator. The amplitude of such beat note is proportional to the field amplitude of the standing wave at the nanosphere position. An example of the beat note amplitude recorded while moving the positioner, nominally along the tweezer axis, is shown in Fig. (\ref{fig_battimento}). We see the Gaussian envelop with a width given by the cavity waist, and a modulation reflecting the sinusoidal standing wave, due to a slight misalignment between the axis of the positioner and the cavity transverse plane. 

The optical cavity is monolitic, built on a 48.8 mm long invar spacing cylinder with a radial hole allowing the inset of the tweezer. The two equal spheric mirrors (nominal curvature radius 25 mm) have measured transmission coefficient of $4.8 \times 10^{-5}$. The cavity linewidth is $\kappa/2\pi = 57$ kHz (Finesse 54000), the efficiency $\eta_{\mathrm{cav}}=\kappa_{\mathrm{out}}/\kappa = 0.41$. 

The overall detection efficiency, taking into account $\eta_{\mathrm{cav}}$, the optical losses, the heterodyne mode-matching ($95\%$) and the photodiodes quantum efficiency, is $\eta = 0.32$.

\begin{figure}
\includegraphics[width=0.8\columnwidth]{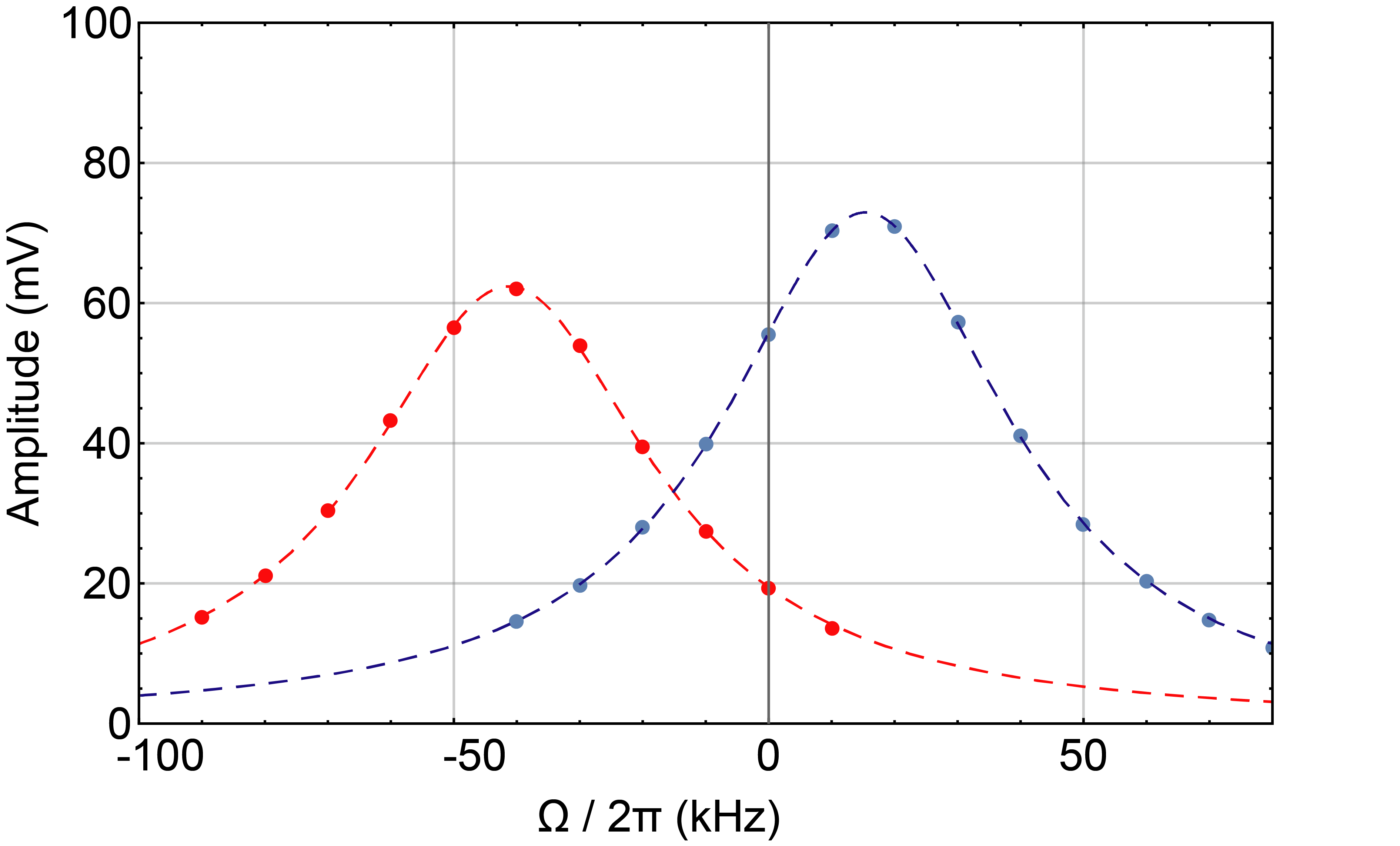}
\caption{Detected intensity of the scanning Nd:YAG2 light transmitted by the cavity, when the reference laser is locked at resonance, as a function of the scanning frequency $\Omega$ (with an offset subtracted). Red (blue) symbols refer to p (s) polarized Nd:YAG2 light when the reference laser is s (p) polarized. Dashed lines show the fitting Lorentzian functions. }
\label{fig_birifrangenza}
\end{figure}

To analyze the transmission lineshape of the empty cavity, we superpose the reference laser beam and a fraction of the beam used for the tweezer, with orthogonal polarizations, before sending them to the cavity input. The reference beam is then locked to a cavity resonance, and the second laser, offset phase-locked to the reference, is scanned over the resonance of the following longitudinal mode using the acousto-optic modulators. The transmission of the second beam is recorded and fitted to a Lorentzian function, as shown in Fig. (\ref{fig_birifrangenza}), deriving the cavity width $\kappa/2\pi= 57.0\pm 0.2$ kHz (the quoted uncertainty reflects the standard error over different measurements). The procedure is repeated swapping the two polarizations. The distance between the two peak centers corresponds to twice the cavity birefringence (that results to be $28.4\pm 0.4$ kHz), their mean value allows to extract the cavity free-spectral-range.  

\subsection*{Noise spectra}

\subsubsection*{Frequency noise}

The frequency noise spectrum of the Nd:YAG lasers, above 10 kHz, is characterized by a structure given by the piezo resonances of the laser crystal, and typically varies between few $10^{-2}$ and 1 Hz$^2$/Hz . The frequency noise has two effects. Inside the optical cavity, it is transformed into intensity noise of the detuned tweezer light scattered by the nanosphere into the cavity mode. Moreover, in the heterodyne detection the same scattered light, transmitted by the cavity, gives a background that partially spoils the heterodyne spectrum. Both effects are strongly reduced in the adopted coherent scattering scheme \cite{delic2020}, since the mean scattered light is nominally null when the nanosphere is positioned on a node of the cavity standing wave, and just a residual due to the nanosphere fluctuations is present.
In our case, such fluctuations are just 5 nm peak-to-peak, and reduced to 0.2 nm by post-selection of $20\%$ of the data. The calculated heating due to the frequency noise is few mHz \cite{delic2020} and it is therefore negligible. However, some residual spectral structures due to the frequency noise remains visible in the heterodyne spectrum above 145 kHz. We have therefore restricted our analysis below this limit, in particular for the spectra at lower pressure.

\subsubsection*{Intensity noise}

The RIN (Relative Intensity Noise) of the tweezer laser, measured in the balanced detection by comparing the sum and the difference signals of the two photodiodes, is $S_{\epsilon} = 2.3 \times 10^{-14}$ Hz$^{-1}$ in the spectral region of interest. The intensity noise spectrum at twice the oscillation frequency causes parametric heating of the nanosphere motion. The heating rate, calculated as $\Gamma_{\mathrm{RIN}} = 0.25 \,\Omega^2 S_{\epsilon}$ \cite{savard1997}, is below 1 Hz, and it is therefore negligible. 

\subsubsection*{Quantum noise calibration}

\begin{figure}
\includegraphics[width=0.8\columnwidth]{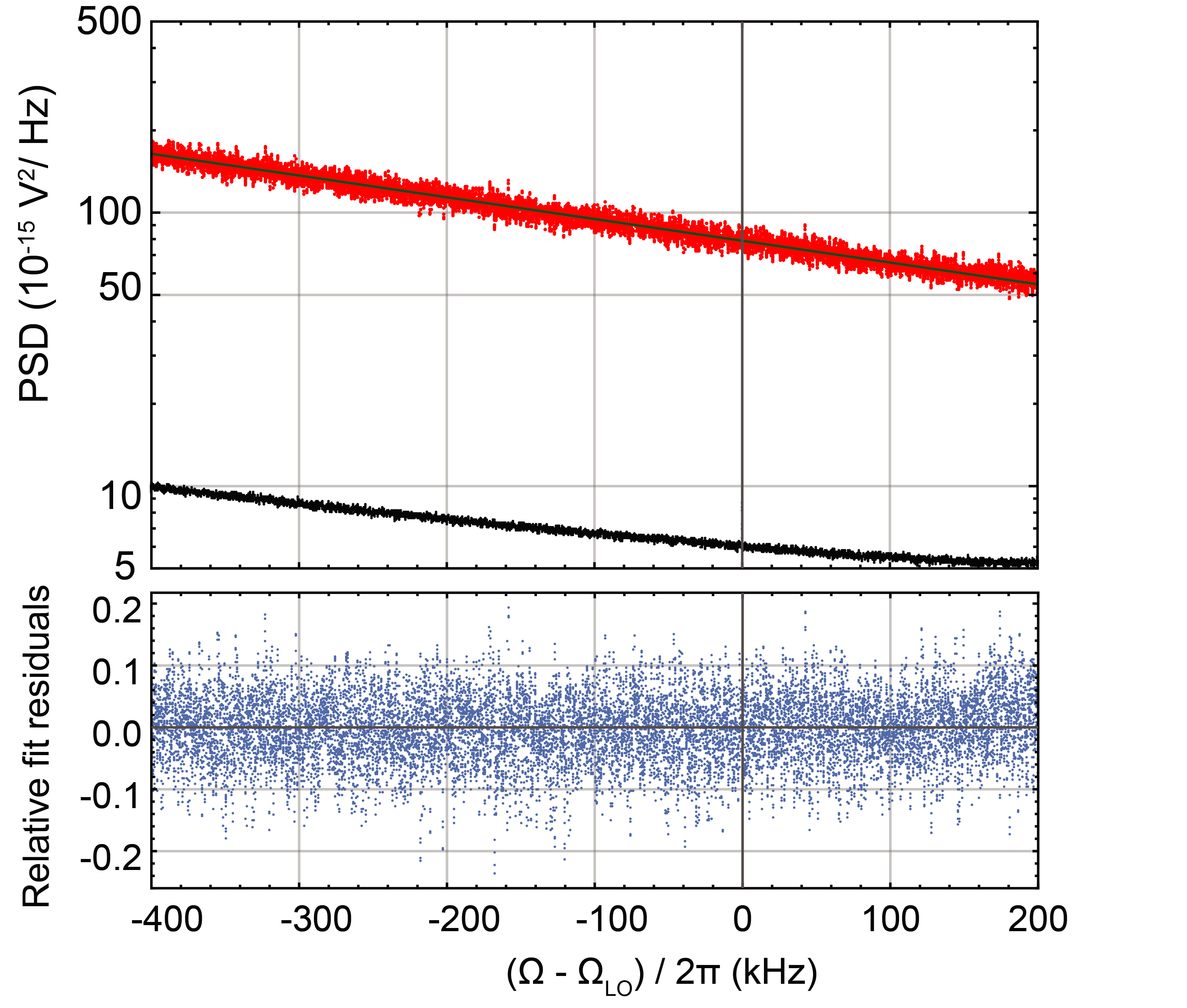}
\caption{Red trace: spectrum of the local oscillator at the output of the balanced detection. The solid line shows the fitting polynomial curve, the fitting relative residuals are shown in the lower panel. Black trace: electronic noise.}
\label{fig_shot}
\end{figure}

In Fig. (\ref{fig_shot}) we show the spectrum recorded by the balanced detection with just the local oscillator, i.e., with vacuum noise as input signal. The local oscillator power is 4.5 mW for all the measurements described in this article. The spectrum nominally corresponds to the reference shot noise, filtered by the overall transfer function of the detectors and the following electronics. Its polynomial fitting function, shown in the figure, is indeed used to normalize the heterodyne spectra to vacuum noise. 

\begin{figure}
\includegraphics[width=0.8\columnwidth]{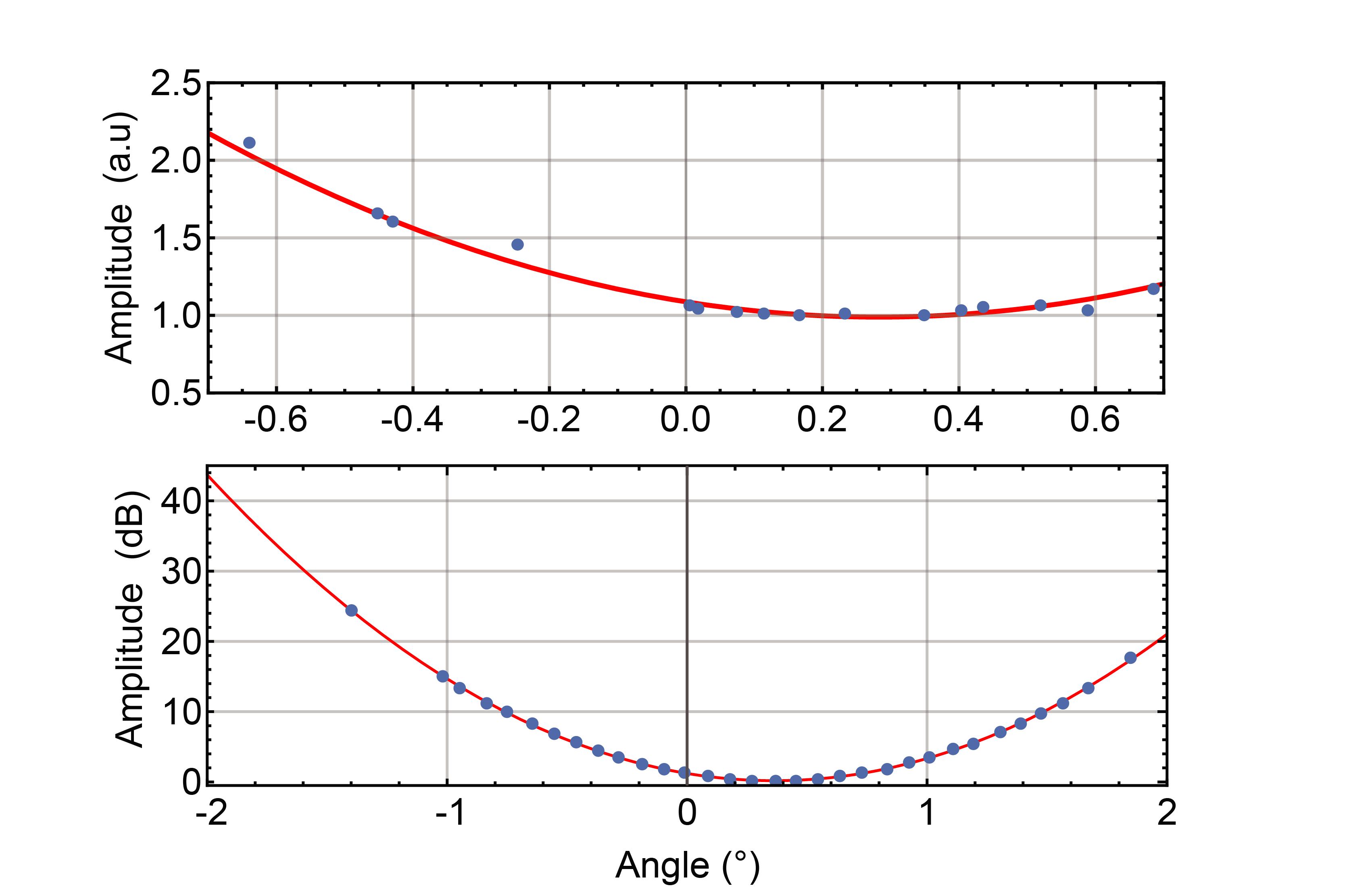}
\caption{Spectral power of the local oscillator, measured at the output of the balanced detection and integrated between 0.7 and 1.5 MHz (upper panel), and amplitude of an intensity modulation peak at 1 MHz (lower panel), as a function of the angle of the input linear polarization in the balanced detection. Both signals are normalized to their minimum value. Red solid lines show the sinusoidal fitting functions.}
\label{fig_rejection}
\end{figure}
\begin{figure}
\includegraphics[width=0.8\columnwidth]{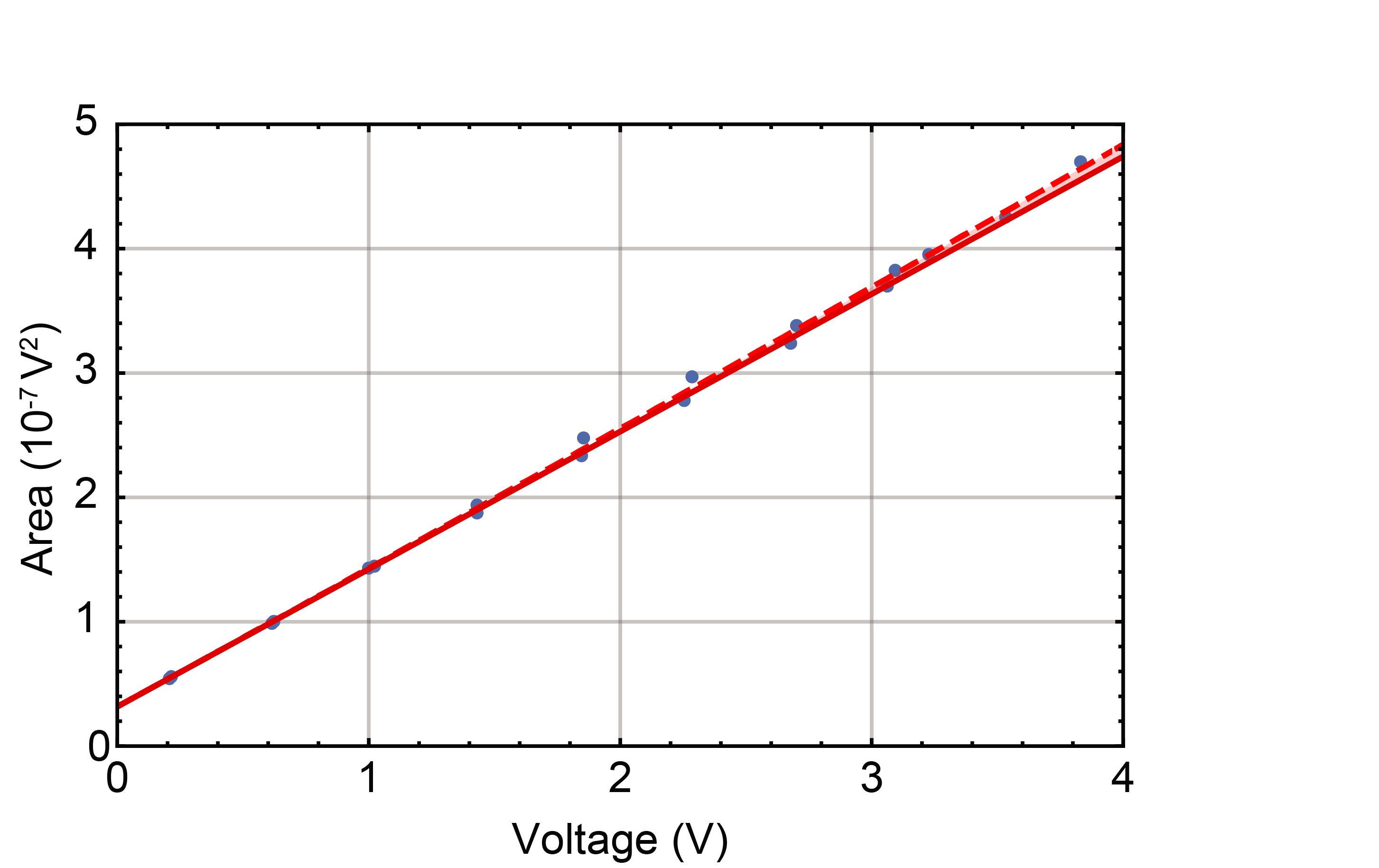}
\caption{Spectral power of the local oscillator, measured at the output of the balanced detection and integrated between 0.7 and 1.5 MHz, as a function of the dc signal in the photodiodes. The red solid line shows the linear interpolation to the data, the dashed line the parabolic fit that indicates an excess noise equivalent to $3\%$ of the shot noise, at the maximum detected power that corresponds to 4.5 mW.}
\label{fig_shotvspower}
\end{figure}

The common mode rejection ratio at 1 MHz and the spectral power integrated between 0.7 and 1.5 MHz are shown in Fig. (\ref{fig_rejection}) as a function of the linear polarization angle at the input of the polarizing beam-splitter of the balanced detection, varied by tuning the angle of the half-wave plate and monitoring the dc output of the balanced detection. This angle is then kept at $0.3^{\circ}\pm 0.1^{\circ}$ during all the measurements. In this range, the common mode rejection is 40 dB. Considering the RIN and the power of the local oscillator, we calculate that the residual amplitude noise at the output of the balanced detection is $\sim 3\%$ of the shot noise level. This is in agreement with the directly measured spectral power, shown in Fig. \ref{fig_shotvspower} for increasing detected power.

\subsection*{Data analysis}

\begin{figure}
\includegraphics[width=0.9\columnwidth]{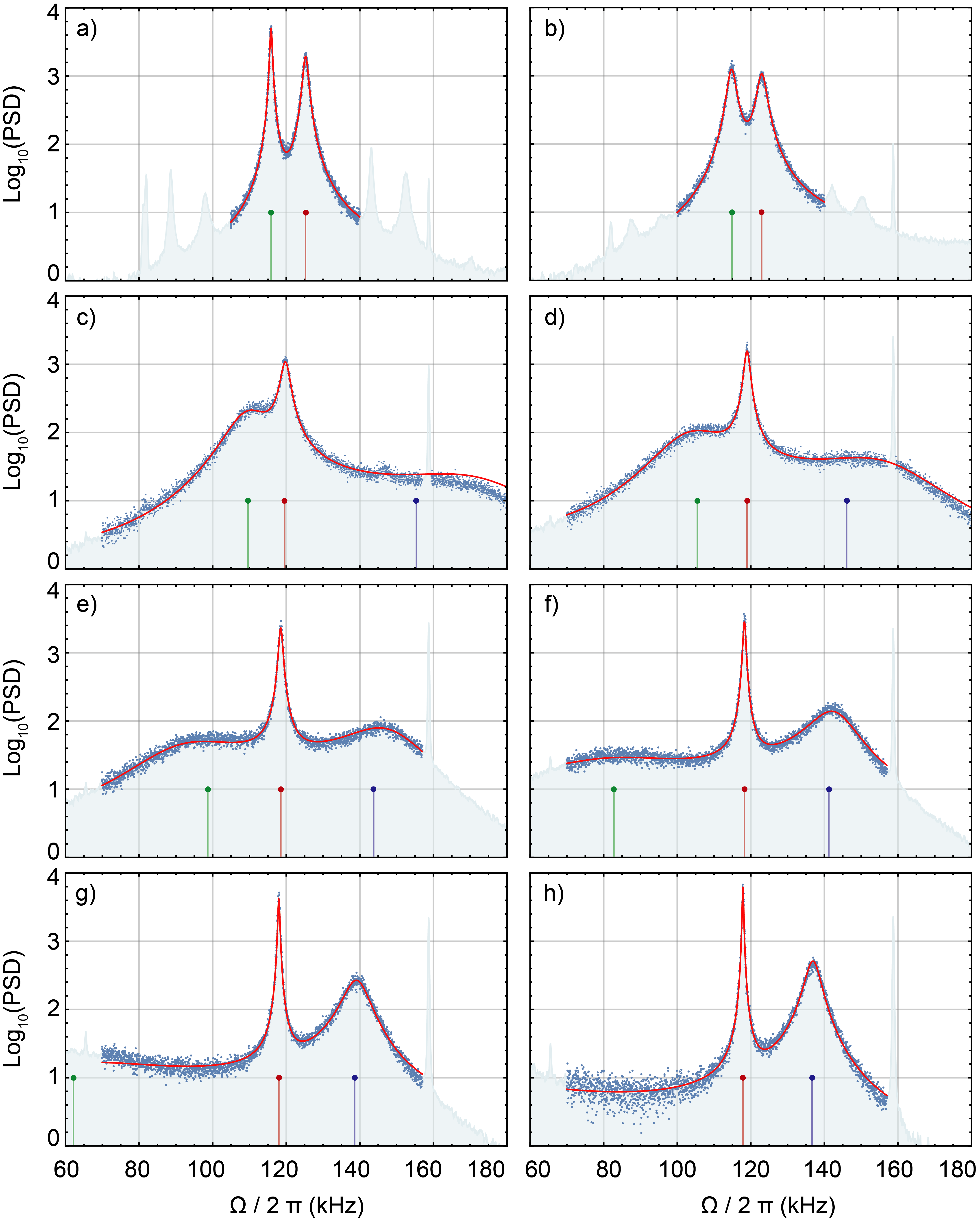}
\caption{High frequency sideband of the heterodyne spectra, for a background pressure of $\sim 6 \times 10^{-3}$ Pa. The detuning of the tweezer light from the cavity resonance is $-\Delta/2\pi = $ 260 kHz (a), 210 kHz (b), 160 kHz (c), 140 kHz (d), 120 kHz (e), 100 kHz (f), 80 kHz (g), 60 kHz (h). Red solid lines show the theoretical spectra, calculated with the following, independently measured, parameters: $\kappa/2\pi = 57$ kHz, $\eta = 0.32$. The mechanical frequencies and optomechanical gains are slightly adjusted for each spectrum, to account for slow variations of the mean nanosphere position and tweezer light power. Their mean values are $\OmegaX/2\pi = 131.6$ kHz, $\OmegaY/2\pi = 117.3$ kHz, $\gX/2\pi = 26.7$ kHz, $\gY/2\pi = 9.4$ kHz.}
\label{fig_spettriSuppl}
\end{figure}

The output signal of the heterodyne balanced detection is low-pass filtered at 1.9 MHz and sampled at 5 MS/s. Spectra are typically obtained with a resolution of 40 Hz, by segmenting the time series into 25 ms long intervals and calculating the Fourier transform. Before averaging, we apply a post-selection of $20\%$ of the intervals, keeping those exhibiting the lowest amplitude of the beat note at 1.1 MHz between the scattered light and the local oscillator, i.e., the signals acquired with the nanosphere closer to the node of the cavity mode. This selection allows to reduce the wings of the beat note peak in the frequency region of interest, as well as to reduce the effect of frequency noise in the spectra.

The spectra shown in Fig. 2 of the main text are displayed again in Fig. \ref{fig_spettriSuppl} in a wider frequency range. Some strong peaks visible in panels (a) and (b), on the sides of the two main peaks associated to the $X$ and $Y$ motion, are due to non-linear effects of the $Z$ motion. The latter is cooled down when decreasing the detuning, due to a small angle between the tweezer transverse plane and the cavity axis. The peak at $\sim 160$ kHz is due to laser frequency noise. The vertical red, green and blue lines mark the positions of the system eigenfrequencies, also shown in Fig. 2b of the main text. 

For large detuning, when the system eigenfrequencies are well separated with respect to the respective withs, each mechanical peak in the output spectrum can be approximated as a Loretzian shape, filtered by the cavity. The area of the peak provides a direct measurement of the decoherence rate $\Gamma_{\mathrm{m}} n_{\mathrm{th}} + \Gamma_{\mathrm{n}}$. We have verified numerically, by comparison with the full model, that this approximated view is valid for $\Delta/2\pi = -260$ kHz, the value that we have used to verify the absence of unmodeled noise sources. The data in Fig. 4a of the main text shows indeed an excellent agreement with the expected behavior at pressures above $10^{-4}$ Pa. Even the deduced behavior of the gas damping (13 Hz $\times P$(Pa)) is in agreement with the calculated one, considering the nominal nanosphere parameters and the $30\%$ absolute accuracy of the pressure gauge. At lower pressure, the peak areas falls below the theoretical curve that just considers gas damping (i.e., $\Gamma_{\mathrm{m}}$ proportional to the pressure) and $\Gamma_{\mathrm{n}}$ due to dipole scattering. A possible explanation can be an error in the measurement of the pressure in high vacuum, when the degassing is close to equilibrium with the pumping rate, due to the distance between the pressure gauge and the cavity. An offset in the pressure of $\sim - 10^{-5}$ Pa would justify the deviation. In this case, the achieved decoherence rate would be around $2\pi \times 30$ kHz, even lower than the $2\pi \times 38$ kHz that we have conservatively estimated in the main text.

\bibliographystyle{apsrev4-2} 
\bibliography{database_v1}

\end{document}